\documentclass[]{aa}
\input{psfig}
\def\eg{{\it e.g.} }
\def\etal{{\it et al.} }
\def\ie{{\it i.e. } }
\def\tomega{\tilde\omega}

\def\dd #1 {{\frac{\partial}{\partial #1}}}
\def\dds{\dd {s} }
\def\Phim{\Phi_{M}}
\def\vt{\vartheta}
\def\tom{\tilde\omega}
\def\cs2{c_{S}^2}
\def\lhs{left-hand side }
\def\rhs{right-hand side }
\def\ltsima{$\; \buildrel < \over \sim \;$}
\def\simlt{\lower.5ex\hbox{\ltsima}}
\def\gtsima{$\;\buildrel>\over\sim\;$}
\def\simgt{\lower.5ex\hbox{\gtsima}}

\authorrunning{Tagger and Pellat}
\titlerunning{Accretion-Ejection Instability}
\begin{document}
\title{An Accretion-Ejection Instability in magnetized disks}

\author{M.Tagger\inst{1} \and R.Pellat\inst{2}}

\offprints{M.~Tagger (tagger@cea.fr)}

\institute{DSM/DAPNIA/Service d'Astrophysique (CNRS URA 2052), CEA 
Saclay, 91191 Gif-sur-Yvette, France \and Centre de Physique 
Th\'eorique, Ecole Polytechnique, Palaiseau, France}

\date{Received date; accepted date}
\thesaurus{02.01.2 - 02.09.01 - 02.13.2 - 08.02.3 - 08.23.3 - 11.10.1}
\maketitle
\begin{abstract}

We present an instability occurring in the inner part of disks 
threaded by a moderately strong vertical (poloidal) magnetic field.  
Its mechanism is such that a spiral density wave in the disk, driven by 
magnetic stresses (rather than self-gravity as in galactic spirals), 
becomes unstable by exchanging angular momentum with a Rossby vortex it 
generates at its corotation radius.  This angular momentum can then 
``leak'' as Alfven waves emitted toward the corona of the disk 
providing, as an element of the accretion process, an energetic source 
for a wind or a jet. As galactic spirals, this instability forms 
low azimuthal wavenumber, standing spiral patterns which might provide 
an explanation for low-frequency QPOs in low-mass X-ray binaries.
\end{abstract}


\section{Introduction}
\label{sec:intro}

The accretion disks of objects such as Active Galactic Nuclei, X-ray 
binaries or Young Stellar Objects are very commonly observed to emit 
jets, whose formation seems to be inherent to the accretion process.  
For instance in YSOs the power of the outflows has been found to be 
correlated with the accretion flux (Cabrit and Andr\'e, 1991 ; Edwards 
\etal, 1993).  In the micro-quasar GRS 1915 recent multi-wavelength 
observations (Mirabel \etal, 1998) show the ejection of plasmoids 
coinciding with the infall of the inner region of the disk toward the 
central black hole.  It is thus considered as a strong point in favor of 
MHD models of jets that, from their very beginning, these models found 
that the jets are very efficient at carrying away the angular momentum 
extracted from the disk, and thus at permitting accretion to proceed.  
This has led to the success of a family of accretion-ejection models for 
these objects ( see \eg Blandford and Paine, 1982; Pelletier and 
Pudritz, 1992).

On the other hand existing models of turbulent accretion in disks 
always involve a transfer of angular momentum which is {\it radial}, 
\ie within the disk: this is built into the ad-hoc model of $\alpha$ 
disks (Shakura and Sunyaev, 1973), where turbulence results in a viscous radial 
diffusion of the angular momentum.  This is also true of specific 
models of disk instabilities (Papaloizou and Pringle, 1984; Tagger 
\etal, 1990,  hereafter paper I; Balbus and Hawley, 1991) which all rely, in one 
form or another, on a radial exchange of angular momentum.  This may 
appear as a contradiction between disk and jet models. We believe 
that it is one of the main reasons why it has always been 
very difficult, though not totally impossible, to develop models 
connecting jet solutions to disk models (Ferreira and Pelletier, 1993; 
Ogilvie and Livio, 1998), in such a manner 
that the angular momentum of accretion is redirected toward the jet.

We present in this paper a new instability mechanism which might help 
solve this contradiction. The instability appears in the 
innermost region of the disk (a few times the inner radius), where 
roughly half of the accretion energy and angular momentum must be 
disposed of. Its physics is such that the angular 
momentum extracted from the disk is not emitted radially but will 
ultimately end up as an Alfven wave traveling to the disk corona, 
where it can thus power the jet. 

This results from the combination of three ingredients:
\begin{itemize}
	\item We have shown in previous work (Paper I) that spiral waves 
	driven by magnetic stresses (rather than by self-gravity as in 
	galactic disks) propagate in disks threaded by a vertical 
	(poloidal) magnetic field.  One can thus consider that the 
	family of spiral density waves in disks concerns disks with an 
	attractive long-range force (self-gravity) as in spiral 
	galaxies, disks with only the local pressure force (giving the 
	Papaloizou-Pringle instability), and disks with a repulsive 
	long-range force, resulting from magnetic stresses.  These waves 
	can be amplified in the region of their corotation radius.  The 
	amplification is strong for self-gravity driven waves, and very 
	weak in the other cases.  However the Papaloizou-Pringle 
	instability concerns only waves with very small wavelengths (of 
	the order of the disk thickness), whereas self-gravity or 
	magnetic stresses move the unstable range to larger scales 
	(small azimuthal wavenumber).

	\item Our result, as well as most of the theoretical work on the 
	propagation and amplification of spirals, was obtained in the 
	classical shearing sheet model, which allows a thorough analysis 
	of the physical processes involved but neglects an important 
	effect: the gradient of vorticity in the equilibrium flow of the 
	disk, and as a consequence the {\it corotation resonance}, whereby 
	the density waves couple to a vortex at their corotation radius 
	\footnote{the radius where the angular phase velocity of the 
	wave equals the angular velocity of the gas, not to be confused 
	with the radius where the gas rotates at the angular velocity of 
	the central object.}; in a paper to be published independently 
	we will present an extension of the shearing sheet model, 
	keeping into account the vorticity gradient in the disk, and 
	thus the physics of the corotation resonance.  It will allow us 
	to analyze in more details the coupling to this vortex, which is 
	in fact a Rossby wave, associated with the vorticity gradient in 
	the disk; the propagation of the Rossby wave is quenched by 
	differential rotation, but this still allows it to exchange 
	energy and momentum with the density wave.  In the present paper 
	we will only use the exact cylindrical geometry, so that the 
	physics of Rossby waves is fully taken into account.\\
	The effect of this resonance is only minor for self-gravity 
	driven spirals, which are very unstable anyway (Pannatoni, 1983).  It 
	does not change much the order of magnitude of the growth rate 
	for the Papaloizou-Pringle instability (Papaloizou and Pringle, 1985; 
	Narayan \etal, 1987).  It has 
	not been considered thus far for the magnetically driven 
	spirals.

	\item On the other hand, we should expect a strong influence in 
	the magnetically driven case: in the Papaloizou-Pringle 
	instability, density waves are evanescent in the corotation 
	region and are thus exponentially small at the corotation 
	radius, where they couple to the vortex.  In the magnetically 
	driven case, density waves create magnetic perturbations which 
	decrease only weakly in the corotation region.  This allows 
	their coupling to the vortex to be much more efficient. More 
	precisely we will find the coupling algebraically, rather than 
	exponentially, small, the small parameter being essentially $m$ 
	(the azimuthal wavenumber) times the ratio of the disk thickness 
	to radius.  We will find this sufficient to get a sizable 
	amplification of the density wave.
	
\end{itemize}
We will show that this instability forms {\it normal modes}, \ie 
standing, exponentially growing patterns. They are localized close to 
the inner radius of the disk, and thus cannot be considered to cause 
accretion in a large portion of it. However in this inner 
region, where roughly half of the accretion energy and angular momentum 
are generated,  they can cause accretion and energize a jet.

One can have two views of this amplification process: we can consider 
it, as we did above, as the amplification of the spiral wave by 
interaction with the Rossby vortex it generates.  But we can also 
concentrate on the evolution of the vortex: differential rotation leads 
it to be sheared away as time evolves, so that an initially leading 
pattern becomes strongly trailing; but at the same time the vortex 
generates a spiral density wave which, after reflection at the inner 
radius of the disk, returns to regenerate it as a leading feature, 
starting a new cycle.

We present two complementary computations of the amplification due to 
this resonance.  The first approach is analytical, in the form of a 
variational principle which allows us to discuss the physics, the sign 
and the order of magnitude of the amplification.  We find in particular 
that the parameter controlling the sign of the resonant effect is not as 
usual the gradient of the specific vorticity in the equilibrium flow, 
but a different quantity involving the magnetic field strength.  This 
first approach allows us to extract and describe in details the physical 
processes involved in the formation of normal modes and in their 
amplification.\\

In a second step we present numerical solutions which confirm this 
behavior, and allow us to exhibit exponentially growing normal modes 
affecting the inner region of the disk. Their growth time is typically 
a fraction of the rotation time of the disk at the corotation radius. 

We derive both of these results in the simple model of an infinitely 
thin disk in vacuum.  In this case the energy and angular momentum 
extracted from the inner region have to stay in the disk: they are 
stored in the vortex at corotation, rather than carried away by a 
density wave travelling outward as in the usual amplification process of 
spiral density waves.  In the Papaloizou-Pringle case, Narayan \etal 
(1987) have argued that the accumulation of angular momentum in the 
vortex might eventually saturate at a finite amplitude, thus ending the 
amplification process. We will briefly discuss how here, if the disk 
were more realistically embedded in a low density corona, the angular 
momentum would be emitted as an Alfven wave emitted vertically along the 
field lines.  We will reserve the detailed discussion and computation of 
this effect to a forthcoming publication.  In our conclusion we will 
discuss possible astrophysical consequences of this new mechanism.

Throughout this paper we will consider {\it normal modes}, \ie 
exponentially growing perturbations characterized by a single frequency 
$\omega$.  As usual in disks with differential rotation, the 
amplification mechanism would also apply to transient shearing 
perturbations, which become normal modes only when a reflexion at the 
inner radius allows them (if they obey an integral phase condition) to 
start a cycle of wave reflection and amplification. The amplification 
process is the same. However it is weak enough that the transient 
amplification of perturbations starting at the noise level is probably 
insufficient to cause accretion; thus only normal modes, whose 
exponential growth allows them to reach large amplitudes, should really 
matter in accretion disks.\\
Curry and Pudritz (1996) have performed a modal analysis of an MHD disk, 
in an incompressible MHD cylinder limit.  Incompressibility did not 
allow them to make a connection with the magnetized spirals of paper I, 
and thus to identify the effects discussed here.  They study in details 
the Alfven resonance ( $\omega-m\Omega(r)=k_{z}v_{A}$, where $k_{z}$ is 
a vertical wavenumber), which is very close to corotation where our main 
effect occurs (corotation is not a singularity in their approximation).  
It is very likely that the Alfven resonance is a form of the coupling to 
vertically propagating Alfven waves which will be briefly discussed 
below.  However in the context of our modes, which are essentially 
constant over a disk scale height, $k_{z}$ looses its meaning and this 
would need to be studied with a more elaborate treatment of the vertical 
dimension, giving the vertical ejection of angular momentum associated 
with our modes.  We defer this to a future publication, and will present 
in Appendix C only a brief sketch of the coupling with Alfven waves far 
from this resonance.
%
\section{The basic model}
\label{sec:model}
\subsection{Thin disk geometry and relation with other instabilities}
\label{subsec:thin}

We restrict ourselves to perturbations of an infinitely thin disk, 
neglecting vertical velocities and variations of the perturbations 
across the disk.  This means that we will not discuss the 
magneto-rotational instability (Velikhov, 1959; Chandrasekhar, 1960; 
Balbus and Hawley, 1991), which relies on the existence of 
a wavelength in the vertical direction.  In a previous work (Tagger 
\etal, 1992) we 
discussed the vertical structure of disk perturbations, and showed that 
there always exists a solution with weak variations across the disk; 
this solution reduces, to lowest order in the disk aspect ratio 
(thickness/radius), to the infinitely thin solution considered here. 

Thus we will totally discard here the vertical structure of the wave
within the disk, which is essential to the magneto-rotational
instability.  There are two reasons for this: one is that these
instabilities are entirely distinct and proceed from very different
mechanisms; the second reason is that the magneto-rotational instability
occurs only when the magnetic pressure is weak, \ie the ratio
$\beta=8\pi p/B^2$ of the thermal to magnetic energy density is larger
than one whereas, as will be seen below, our instability is significant
only when $\beta$ becomes of the order of one, or weaker.  We will thus
consider disks where the magnetic pressure is strong, in the sense that
$\beta\simlt 1$, but still much weaker than the gravitational energy so
that magnetic support of the disk remains weak.  When magnetic support
becomes strong our instability becomes the interchange mode studied by
Spruit and coworkers (Spruit and Taam, 1990; Spruit \etal, 1995) in
disks with and without differential rotation.  Indeed our basic
assumptions are essentially the same as theirs.  In an approximate
analysis of the effect of differential rotation Spruit \etal (1995)
found that it makes the amplification of the interchange mode only
transient, and negligible unless the magnetic field is very strong.  In
fact the vortex part of our instability may be viewed in some respects
as an interchange, since the vortex effectively exchanges flux tubes
across the corotation radius, and the amplification relies on the radial
gradients of $B_{0}$ and $\Sigma$ (the surface density) -- although in a
different form than that found by Spruit \etal.  The coupling between
the vortex and density waves, which allows the formation of
exponentially growing normal modes, allows our instability to act
already at $\beta\sim 1$, \ie at much lower magnetic pressure than in
their case.  A more detailed comparison is difficult, because we use a
slightly different model: Spruit \etal consider magnetic fields
supported by equilibrium currents only in the disk, resulting in a jump
of $B_{r}$ across the disk, whereas for simplicity we assume that the
magnetic field gradient at equilibrium is only due to external currents
(\eg ring currents inside the inner disk radius), so that the
equilibrium field is purely vertical across the disk.

Our justification for this is that we are interested in a regime where 
the interchange instability should be negligible.  On the other hand in 
more recent work, Stehle and Spruit (1998) found in non-linear 
simulations strong instabilities even in situations which should be 
stable to the interchange mode.  It is possible that this is a 
manifestation of the instability we present here (in particular the 
density and magnetic field profiles they use should give instability, 
according to our criterion).  But the growth rates they find are rather 
large, and the modes have a more global structure than we would expect.  
Although no definite conclusion can be reached at this stage, we rather 
believe that their result is due to the reflective boundary condition 
they use at large $r$: in the Papaloizou-Pringle case (Narayan \etal, 
1987) and presumably also in the magnetic case, this leads to a strong 
form of the spiral instability, due to the formation of a double cavity 
between the corotation radius and the inner and outer radii of the disk.  
This boundary condition, often used in disk theories and simulations, 
can thus be quite misleading, although one might look for astrophysical 
justifications in specific contexts, if some mechanism can act to 
truncate the disk at large radii.  Otherwise a radiation condition at 
the outer radius must be used, as we do here.  Simulations with 
different boundary conditions, or increased outer radius, would then be 
necessary to fully understand their result.  

Our simple model of a disk in vacuum also means that we cannot fully 
justify, in the limits of the present paper, our claim that our 
instability may eventually redirect toward the disk corona the angular 
momentum it extracts from the disk.  This can however be easily 
understood if one remembers that the disk is threaded by magnetic field 
lines, along which Alfven waves can propagate as soon as there is any 
small plasma density in the corona of the disk; the vortex at the 
corotation radius moves the footpoints of the magnetic field lines, and 
this motion (of torsional nature, whereas the motion associated with the 
spiral is essentially compressional) will act as an initial condition 
for the emission of Alfven waves.  We can thus consider that, although 
this is not contained in the present description where the disk is in 
vacuum, the energy and angular momentum deposited in the vortex by the 
instability will end up in these Alfven waves propagating in the corona.  
In Appendix C we show how this occurs in the region where 
density waves propagate, in the WKB approximation.  The full 
computation, valid in the corotation region, is much more complex and we 
defer it to a future publication.

\subsection{The magnetic field}
\label{subsec:mag}

We consider a disk threaded by a vertical magnetic field.  For 
simplicity we discard radial or azimuthal (toroidal) components of the 
equilibrium magnetic field, and assign its radial dependence to external 
currents\footnote{This choice is made in order to trim down our model 
to the minimal necessary physics, for the sake of clarity in this paper 
where we wish to present and discuss the basic physics of our 
instability.  Taking into account equilibrium currents in the disk, 
which cause a finite inclination of the field lines at its surface, is 
straightforward, and preliminary results show that they increase 
significantly the growth rate of the instability.}.  Thus at 
equilibrium the disk is characterized by its surface density $\Sigma(r)$ 
and magnetic field $\vec B=B_{0}(r)\vec e_{z}$.

The disk lies in vacuum: then the absence of currents allows us to 
describe the perturbed magnetic field outside the disk by a magnetic 
potential:

\begin{equation}
\label{eqn:Phim}
{\vec B }=-{\rm sign}(z)\vec\nabla\Phim
\end{equation}
with 
\begin{equation}
\label{eqn:lapphi}
\Delta\Phim=0.
\end{equation}
We work in cylindrical coordinates 
($r,\ \vt, \ z$).  The sign of $z$ in Eq.~(\ref{eqn:Phim}) 
ensures that, if $\Phim$ is even in $z$, the perturbed vertical field 
is also even, while its horizontal ($r$ and $\vt$) components are odd.  
From their change across the disk we get the currents:
\begin{eqnarray*}
    j_{r} & = & 2\dd {\vt} \Phim\delta(z)\\
    j_{\vt} & = & -2\dd r \Phim\delta(z)
\end{eqnarray*}
and the horizontal magnetic stresses acting on the disk:
\begin{eqnarray*}
    F_{M}^r & = &  -2B_{0}\dd r \Phim\delta(z)\\
    F_{M}^\vt & = & - 2B_{0}\dd{\vt} \Phim\delta(z)
\end{eqnarray*}
On the other hand the vertical component of the field:

\[B_{z} =  -{\rm sign}(z)\dd z \Phim \]
gives at the disk:

\[\frac{\partial^2}{\partial z^2}\Phim=-2B_{z}^{D}\delta(z)
\]
where $B_{z}^{D}$ is the perturbed field in the disk. 
Thus throughout space we can write:

\begin{equation}
\label{eqn:PoissonB}
\Delta\Phim=-2B_{z}^{D}\delta(z)
\end{equation}
This can be compared with the Poisson equation giving the perturbed 
gravitational potential in a self-gravitating disk:
\begin{equation}
\label{eqn:PoissonG}
\Delta\Phi=4\pi G \sigma \delta(z)
\end{equation}
where $\sigma$ is the perturbed surface density.  We can thus compute 
the magnetic potential by the classical Poisson Kernel (Binney and 
Tremaine, 1987): assuming perturbations varying as exp($im\vt$) we have:

\begin{equation}
\label{eqn:kernel}
\Phim(r)=\int dr' K_{m}(\frac{r'}{r})B_{z}^{D}(r')
\end{equation}
where

\begin{equation}
\label{eqn:Kernel}
K_{m}(x)=\frac{x}{\pi}\int_{0}^{\pi}\frac{d\vt\cos 
m\vt}{(x^2-2x\cos\vt+1)^{1/2}}
\end{equation}

\subsection{Perturbation equations}
\label{subsec:pert}
We write the linearized Euler equations for perturbations varying as 
$e^{i(m\vt-\omega t)}$:
\begin{eqnarray*}
    -i\tom v_{r}-2\Omega v_{\vt} & = & \frac {F_{r}}{\Sigma}\\
    -i\tom v_{\vt}+W v_{r} & = & \frac {F_{\vt}}{\Sigma}
\end{eqnarray*}
where the forces in the right-hand sides comprise pressure and 
magnetic stresses.  Here $\tom=\omega-m\Omega(r)$, $\Omega$ is the 
rotation frequency of the equilibrium flow, and $W$ is known in the 
context of galactic disks as Oort's second constant\footnote{we 
will not use here the more classical notation $B$, in order to avoid 
confusion with the magnetic field}:

\[W=\frac{\kappa^2}{2\Omega}=\frac 1 {r}\frac d {dr}(r^2\Omega)\] 
$\kappa$ is the epicyclic 
frequency, and it is worth noting already at this point that $W$ is 
the vertical component of the vorticity of the equilibrium flow.

We find it more simple to change variables to: 
\begin{eqnarray*}
     s&=\ln r\\
     U&=rv_{r}\\
     V&=rv_{\vt}
\end{eqnarray*}
We get:
\begin{eqnarray}
    -i\tom U-2\Omega V & = & -\cs2 \dds {h} -2\frac{B_{0}}{\Sigma}\dds \Phim
    \label{eqn:Eulu}  \\
    -i\tom V+WU & = & -im\cs2 h -2im\frac{B_{0}}{\Sigma}\Phim
    \label{eqn:Eulv}
\end{eqnarray}
where we have defined

\[h=\frac{\sigma}{\Sigma},\] $\sigma$ is the perturbed density, and 
$c_{S}$ is the sound speed.  For simplicity we have assumed an 
isothermal equation of state, and we will consider throughout this paper 
that $c_{S}$ is constant.  We also write the continuity equation:

\begin{equation}
    -i\tom r^2\sigma=-\dds (\Sigma U) -im\Sigma V
    \label{eqn:contin}
\end{equation}
Our  system of equations is closed by the vertical component of the 
induction equation :

\begin{equation}
    -i\tom r^2B_{z}^D=-\dds (B_{0} U) -imB_{0} V
    \label{eqn:contB}
\end{equation}
which can be usefully compared with Eq.~(\ref{eqn:contin}): they 
correspond respectively to the conservation of magnetic flux and total 
density in a flux tube.  Let us now consider 
Eqs.~(\ref{eqn:contin}-\ref{eqn:contB}) and 
(\ref{eqn:PoissonG}-\ref{eqn:PoissonB}).  The parallel between $\sigma$ 
and $B_{z}$, $\Phi$ and $\Phim$, shows that, as explained in paper I, 
the magnetic field appears in the mathematical form of a negative 
self-gravity, \ie a long-range force whose action is repulsive (the 
magnetic field stiffening the field lines and thus resisting, rather 
than favoring, the formation of overdensities).  We will hereafter, for 
simplicity, assume that the self-gravity of the disk is negligible but 
its inclusion in our description to discuss more massive disks would be 
straightforward.

\section{Corotation and Amplification}
\label{sec:analyt}
\subsection{A variational principle}
\label{subsec:variat}
In this section we derive a variational principle which will allow us to 
analyze the effect of the corotation resonance, and also to discuss the 
boundary conditions to be used.  As usual with variational formulations, 
this permits a more rigorous and general result than in previous 
discussions of both the resonance and the boundary conditions.  In order 
to minimize the complexity, we present here only a simplified form 
retaining only magnetic stresses; the 
full computation, including pressure forces, is given in Appendix~A.

We start from Eqs.~(\ref{eqn:Eulu}-\ref{eqn:Eulv}), which give:
\begin{eqnarray}
    U  =  \frac 1 L \biggl\{
    -2i\tom\frac{B_{0}}{\Sigma}\dd s \Phim 
    +4im\Omega\frac{B_{0}}{\Sigma}\Phim\biggr\}
     \label{eqn:UL}\\
     V  =  \frac 1 L \biggl\{
    -2W\frac{B_{0}}{\Sigma}\dd {s} \Phim
    +2m\tom\frac{B_{0}}{\Sigma}\Phim\biggr\}
     \label{eqn:VL}
\end{eqnarray}
where 
\[L=\tom^2-\kappa^2\]
Substituting into Eq.~(\ref{eqn:contB}), and using:
\[W=2\Omega+\Omega',\ \ \ \dd {s} \tom=-m\Omega'\]
where the prime denotes the derivative with respect to $s$, we get:
\begin{eqnarray}
    i\tom r^2&\displaystyle{B_{z}^{D}  =  -i\tom  {\dd {s} }\biggl\{ 
    2\frac{B_{0}^2}{\Sigma L}\dd{s} \Phi_{M} 
     \biggr\} }\ \ \ \ \ \ \ \ \ \ \ \ \ \ \ \ \ \ \ \nonumber
     \\
     &\displaystyle{+\Phi_{M}\left[\frac{\partial}{\partial s} 
     \left(\frac{4im\Omega 
     B_{0}^2}{\Sigma L}\right)+ 
     2i m^2 \frac{B_{0}^2}{\Sigma}\frac{\tom}{L}\right]}
\label{eqn:Bz}
\end{eqnarray}
After some algebra we write this as:
\begin{eqnarray}
& r^2B_{z}^D=-\displaystyle{\dd {s} \,\left( \frac{2}{L}\frac{B_{0}^2} 
{\Sigma}\, \dd {s} \Phi_{M}\right)}\ \ \ \ \ \ \ \ \ \ \ \ \ \ \ \ \ \ \ 
\ \ \ \ \ \ \ \ \ \ \ \ \ \ \ \nonumber \\
 &\nonumber\\
&\displaystyle{+\frac{2B_{0}^2}{\Sigma L^2}\Phi_{M} 
\biggl\{m^2\left(L+4\Omega\Omega'\right)}\ \ \ \ \ \ \ \ \ \ \ \ \ 
\nonumber \\
 &\nonumber\\
    &\!\!\!\!\!\!\!\!\!\!\!\!\!\!\!\!\displaystyle{+2m\Omega\biggl[ 
    \displaystyle{\tom\dd {s} \left(\ln\frac{\Omega 
    B_{0}^2}{\Sigma}\right)}+2\frac{\Omega 
    W}{\tom}\dd {s} 
    \left(\ln\frac{W\Sigma}{B_{0}^2}\right)\ \biggr]\ \biggr\}}\\
\label{eqn:Bzd}
 &\nonumber
\end{eqnarray}
We multiply both sides of this equation by $\Phi_{M}^*$ (where the star 
indicates the complex conjugate) and integrate over $s$ between 
boundaries $s_{min}$ and $s_{max}$. After an integration by parts we get 
a quadratic form:\\
\begin{eqnarray*}
&\displaystyle{\int_{s_{min}}^{s_{max}}ds }&\displaystyle{
\biggl\{r^2(s)\Phi_{M}^*(s)B_{z}^{D}(s)} \ \ \ \ \ \ \ \ \ \ \ \ \ \ \ \ 
\ \ \ \ \ \ \ \ \ \ \ \ \ \ \ \ \ \ \ \ \ \ \ \ \ \ \ \ \ \ \ \ \ \ \ \ 
\ \ \ \ \ \ \ \ \ \ \ \ \ \ \ \\
& &\\
&&-\displaystyle{\frac{2}{L}\frac{B_{0}^2}{\Sigma}\left|\dd {s} 
\Phi_{M}\right|^2}\ \ \ \ \ \ \ \ \ \ \ \ \ \ \ \ \ \ \ \ \ \ \ \ \ \ \ 
\ \ \ \ \ \ \ \ \ \ \ \ \ \ \ \ \ \ \ \ \ \ \ \ \ \ \ \ \ \ \ \ \ \ \ \ 
\ \ \ \ \\
& &\\ 
&&-\displaystyle{\frac{2B_{0}^2}{\Sigma L^2}\left|\Phi_{M}\right|^2 
\biggl\{m^2\left(L+4\Omega\Omega'\right)}\ \ \ \ \ \ \ \ \ \ \ \ \ \ \ \ 
\ \ \ \ \ \ \ \ \ \ \ \ \ \ \ \ \ \ \ \ \ \ \ \ \ \ \ \ \ \ \ \ \ \ \ \ 
\ \ \ \ \ \ \ \ \ \ \ \ \ \ \ \ \ \ \ \ \ \ \ \ \ \ \ \ \\
& &\\
     && \!\!\!\!\!\!\!\!\!\!\!\!\!\!\!\!\!\!\!\!\!\!\!\!\!  
     \displaystyle{+2m\Omega\biggl[ \tom\dds \left(\ln\frac{\Omega 
     B_{0}^2}{\Sigma}\right)+2\frac{\Omega W}{\tom}\dds 
     \left(\ln\frac{W\Sigma}{B_{0}^2}\right)\ \biggr]\ \biggr\}\ 
     \biggr\}}\\
\end{eqnarray*}
\begin{equation}
    =-\displaystyle{\biggl[\frac{2}{L}\frac{B_{0}^2}{\Sigma}\Phi_{M}^{*}\dd 
{s} \Phi_{M}\biggr]_{s_{min}}^{s_{max}}}
\label{eqn:varprinc}
\end{equation}

Let us consider the first term in the integral, in the \lhs of this 
equation: using (\ref{eqn:kernel}) we can write it as:
\[\int ds\,r^2 B_{z}^D(s)\int ds'\,
r'K_{m}\left(\frac{r'}{r}\right)B_{z}^{D*}(s')\]
Using the property that:
\[K_{m}\left(\frac{1}{x}\right)=\frac{1}{x}K_{m}(x)\]
we see that this expression is hermitian (symmetric to the exchange of $B_{z}^D$ 
and $B_{z}^{D*}$). 

Let us now consider the other terms in the \lhs of 
Eq.~(\ref{eqn:varprinc}): they have resonant denominators at $L=0$ (the 
Lindblad resonances, $\tom=\pm \kappa$) and at $\tom=0$ (the corotation 
radius).  We will admit without demonstration the common result 
(obtained by a Frobenius expansion) that the solutions are regular at 
Lindblad resonances (Narayan \etal, 1987), but not at corotation where 
the waves can exchange energy with the flow.  This resonance has been 
discussed by Pannatoni (1983) in the context of galactic spiral waves, 
and by Papaloizou and Pringle (1985) and Narayan \etal (1987) in the 
context of the Papaloizou-Pringle instability.  Its effect on the 
magnetically driven spiral density waves is the main object of this 
paper.

The corotation resonance appears only in the last term in the integral, on the 
\lhs of Eq.~(\ref{eqn:varprinc}).  As will be seen later, 
it contributes an imaginary term to the integral and thus makes it 
non-hermitian.  We will consider, as in the above-mentioned treatments, 
that this term is small (this can be obtained by reducing the derivative 
of $B_{0}^2/W\Sigma$), so that we can treat it as a perturbation.  
Then to lowest order, all the terms in the \lhs are Hermitian, 
and the \rhs is a boundary term (containing the physics 
associated with boundary conditions, as will be discussed below).  This 
means that Eq.~(\ref{eqn:varprinc}) is self-adjoint, so that it 
forms a variational principle (Morse and Feschbach, 1953).  Appendix A shows 
that, as could be expected, this property is still valid when we retain 
pressure forces.

For readers not familiar with variational principles, we will only 
mention here their main advantage, namely that they allow a rigorous 
treatment of perturbation problems: let us note the variational form as: 
\[ \left<\xi^*|{\cal L}(\omega)|\xi\right>=0\] 
where ${\cal L}={\cal L}_{0}+\varepsilon{\cal L}_{1}$, $\varepsilon$ is 
a small parameter and $\xi$ is the vector formed from the components 
(velocities, potential, density) of the perturbation.  We assume that we 
know ($\omega_{0},\ \xi_{0}$), an eigenvalue and eigenvector (the 
frequency and spatial dependence of the solution) of the unperturbed 
problem ${\cal L}_{0}$:
\[{\cal L}_{0}(\omega_{0})\xi_{0}=0\]
and look for 
solutions ($\omega_{0}+\delta\omega,\ \xi_{0}+\delta\xi$) of the 
perturbed problem, to first-order in $\varepsilon$.  The hermiticity of 
${\cal L}_{0}$ gives
\[\left<\xi_{0}^*|{\cal L}_{0}(\omega_{0})|\delta\xi\right>=
\left<\delta\xi^*|{\cal L}_{0}(\omega_{0})|\xi_{0}\right>=0\]
so that $\delta\omega$ can be obtained to first order without need to 
compute $\delta\xi$, making the task relatively simple and more 
straightforward than the ``direct'' approach used \eg by NGG. One gets 
simply:
\begin{equation}
    \left<\xi_{0}^*\left|\frac{\partial {\cal 
    L}_{0}}{\partial\omega}\right|\xi_{0}\right> 
    \delta\omega=-\varepsilon<\xi_{0}^*|{\cal 
    L}_{1}(\omega_{0})|\xi_{0}>
\label{eqn:pert}
\end{equation}
In section \ref{subsec:WKB} we discuss the main properties of the 
unperturbed solutions. Since this neglects the corotation resonance, 
they are essentially the properties of magnetically-driven spiral 
waves, which had been obtained in the shearing-sheet approximation in 
Paper I. In section \ref{subsec:Reson} we will perturb them by the 
corotation resonance term. The boundary conditions, appearing in the 
\rhs of Eq.~(\ref{eqn:varprinc}), are discussed 
separately in Appendix B and implemented in the numerical solution 
presented in section (\ref{sec:Num}).
\subsection{Waves and modes}
\label{subsec:WKB}
In this sub-section we discuss waves and modes, as they can be obtained 
from a WKB approximation in the unperturbed state (without the 
corotation resonance).  The WKB approximation applies for spiral waves 
when the radial wavenumber is much larger than the azimuthal one (giving 
tightly wound spirals).  Then the first two terms, on the \lhs of 
Eq.~(\ref{eqn:varprinc}) are larger than the other ones, which are 
proportional to $m$.  We look for a local radial wavenumber $k$:
\[\dds \Phi_{M}\approx i k(s) \Phi_{M}\]
and use the relation between $\Phi_{M}$ and $B_{z}^D$:
\[B_{z}^D=-\left[\dd z \Phi_{M}\right]_{z=0^+}\]
together with Eq.~(\ref{eqn:lapphi}) which gives:
\[\dd z \Phi_{M}=-\frac{|k|}{r} \Phi_{M}\]
where the minus sign comes from the constraint that $\Phi_{M}$ 
decreases at $z\rightarrow\infty$.  Then from the first two terms in 
Eq.~(\ref{eqn:varprinc}) we get the dispersion relation:
\begin{equation}
    \tom^2=\kappa^2 +\frac{2B_{0}^2}{\Sigma} \frac{|k|}{r}
    +\frac{k^2}{r^2}\cs2
    \label{eqn:reldisp}
\end{equation}
where for completeness we have added the pressure term, derived from 
the full variational principle in Appendix A. This is the dispersion 
relation derived in paper I in the shearing sheet approximation. It 
is identical to the dispersion relation of self-gravity driven spiral 
density waves, with the magnetic term acting as a negative 
self-gravity.

From this dispersion relation one gets the basic properties of these 
waves: one can have both {\it leading} ($k<0$) and {\it trailing} 
($k>0$) waves, resulting in spiral structures wound in opposite 
directions.  Their group velocities, on both sides of corotation, are 
respectively toward and away from the corotation radius (where $\tom$ 
vanishes).

The waves propagate only where Eq.~(\ref{eqn:reldisp}) has a 
positive root for $|k|$, \ie for $\tom^2-\kappa^2>0$ (beyond the 
Lindblad resonances).  In the corotation region they are evanescent.  
Thus a leading wave traveling from the center toward corotation will be 
reflected at the Inner Lindblad Resonance (ILR) as a trailing wave 
traveling back toward the center.  If the boundary condition at or near 
the center is, as usually considered, such that this trailing wave is 
reflected again as a leading wave, one can establish a cavity between 
the center and the ILR; for a discrete set of frequencies, given by an 
integral phase condition, the resulting leading wave has the same phase 
as the initial one, so that a standing pattern can be established.  
These patterns are known as normal modes of the system.

Another property of the waves is that they have a {\it negative energy} 
inside corotation, because they rotate slower than the gas, and positive 
energy outside.  If the waves can tunnel through the forbidden band at 
corotation, they are amplified: let us explain this by considering a 
leading wave of energy -1 traveling from the center to the ILR, 
resulting by tunnel effect in the emission beyond the Outer Lindblad 
Resonance (OLR) of a trailing wave of energy $+\varepsilon$, unspecified 
but positive.  Then the trailing wave reflected to the center must have 
energy $-1-\varepsilon$, \ie a higher amplitude than the incoming wave: 
as it is reflected at the ILR, the wave is amplified.

In the case of the Papaloizou-Pringle instability, where the only force 
comes from the pressure (which acts only locally), $\varepsilon$ is 
exponentially small: this is natural since the waves are evanescent in 
the forbidden band, as usual with tunnel effects, so that the energy of 
the transmitted wave is proportional to the exponential of a large 
negative quantity, of the order of $-r\Omega/c_{S}$.  However we have 
shown in PTS and paper I that the long-range nature of the forces change 
this result in the case of self-gravity driven or magnetically driven 
spirals.  This appears mathematically, in the shearing sheet model, as 
the contribution of a branch cut in the complex-$k$ space, related to 
the spatial behavior of the potentials.  The reason is in fact quite 
simple: consider a density perturbation in the cavity between the center 
and the ILR. The tunnel effect allows it to act beyond the forbidden 
band, in the same manner as in the Papaloizou-Pringle case.  However, 
this density perturbation creates a potential which is {\it not} 
evanescent in the forbidden band!  As usual for an $m$-polar potential, 
it varies as $r^{-m}$, so that it decreases only algebraically across 
the corotation region.  This radial dependence corresponds to the fact 
that the branch cut found in $k$-space starts from singular points at 
$k=im$, or $k_{x}=ik_{y}$ in the cartesian geometry of the shearing 
sheet.

The result of this anomalous tunneling is that, for self-gravity driven 
spirals, $\varepsilon$ can be very large, rather than exponentially 
small as in the Papaloizou-Pringle case.  In the magnetically driven 
case, we found in Paper I that maximal amplification is shifted to smaller 
azimuthal wavenumber, but remains small.  We show in the next 
sub-section that the corotation resonance can make it much larger.
\subsection{Resonant effect at corotation}
\label{subsec:Reson}
We can now turn to the perturbative treatment of the corotation resonance. 
For simplicity we will neglect boundary terms (the \rhs of 
Eq.~(\ref{eqn:varprinc})), treated separately in Appendix B. We 
thus rewrite Eq.~(\ref{eqn:varprinc}) as:

\begin{eqnarray}
\displaystyle{\int_{s_{min}}^{s_{max}}\!\!\!\!ds }&\displaystyle{
\biggl\{r^2(s)\Phi_{M}^*(s)}\displaystyle{B_{z}^{D}(s) }&\ \ \ \ \ \ \ \ \ \ \ \ \ \ \ \ 
\ \ \ \ \ \ \ \ \ \ \nonumber\\
 &\nonumber\\
&-\displaystyle{\frac{2}{L}\frac{B_{0}^2}{\Sigma}\left|\dd {s} 
\Phi_{M}\right|^2}&-\displaystyle{\frac{2B_{0}^2}{\Sigma L^2}\left|\Phi_{M}\right|^2 
{\cal O}(m)\biggr\}}\nonumber\\
&\nonumber\\
={\int_{s_{min}}^{s_{max}}\!\!\!\!\!\!\!\!ds }&\displaystyle{ 
\frac{8m\Omega^2 W}{\tom}\frac{B_{0}^2}{\Sigma 
L^2}}&\displaystyle{\left|\Phi_{M}\right|^2 \dds}\displaystyle{ 
\left(\ln\frac{W\Sigma}{B_{0}^2}\right)}\label{eqn:perturb1}\\
&\nonumber
\end{eqnarray}
where ${\cal O}(m)$ contains terms we neglected in the WKB 
approximation; we will also neglect them here as in the ``cavity'' where 
the waves propagate (so that $\Phi_{M}$ is large) these terms are small.  
To lowest order (\ie neglecting the \rhs) this equation gives a discrete 
spectrum of normal modes, discussed in the previous sub-section, which 
are neither amplified nor damped (since the operator is hermitian, 
$\omega$ is real).

To the next order we perturb these solutions by the resonant term in the 
\rhs, using Eq.~(\ref{eqn:pert}).  We are only interested in the 
imaginary part of the integral in the \rhs, since it will give us the 
growth rate resulting from the resonance.  This is done in the classical 
way used to study Landau damping in plasmas, or similar fluid problems 
(Lin, 1955, Narayan \etal, 1987), and for completeness we will sketch 
here its justification.
\begin{figure}
\psfig{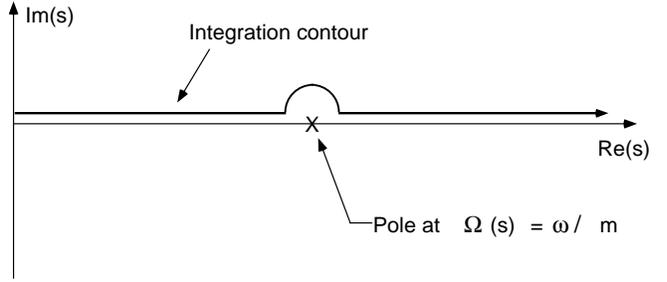}
\caption{\label {fig:contour} \scriptsize
The integration path for the integral in Eq.~(\ref{eqn:perturb1}). 
The contour lies on the real axis (giving a real Principal Part), 
except for a clockwise semi-circle around the corotation pole.
}\end{figure}
One must remember that, since the modes we are looking for are 
exponentially growing or decreasing, the use of a frequency $\omega$ 
cannot be justified by a Fourier but by a Laplace transform; its 
inversion is obtained by an integral on a path in the complex-$\omega$ 
plane, which must lie above (\ie at larger imaginary part than) any 
singularity.  In Eq.~(\ref{eqn:perturb1}) the singularity is at $s$ such 
that $\omega=m\Omega(s)$, and we must compute the integral in the \rhs 
as if $\omega$ had a positive real part.  One finds easily that this 
means that the pole $s_{cor}$ lies below the real-$s$ axis, so that the 
integration can be done on the contour shown in Fig.~\ref{fig:contour}: 
away from the pole the path stays on the real-$s$ axis, so that its 
contribution is the principal part of the integral: this is essentially 
real, so that it only contributes to a small change of the real part of 
$\omega$.

The rest of the integration contour is a half-circle around the pole, 
described clockwise because the pole (when the inverse Laplace transform 
is properly defined, \ie when $\omega$ has a positive imaginary part) is 
below the real axis: its contribution to the integral is thus, from 
Cauchy's theorem, $-i\pi$ times the residue of the integrand at the pole.  
This gives:
\begin{eqnarray*}
    \displaystyle{\delta\omega}&\displaystyle{ \int_{s_{min}}^{s_{max}}
    ds\,\,\,
    2\frac{B_{0}^2}{\Sigma} \left|\dd {s} \Phi_{M}\right|^2
     \dd {\omega} \left(\frac {1} {L}\right) }&\\
    &&\\
    &\displaystyle{=\frac{i\pi}{\partial \tom / \partial 
    s}\biggl[8m\frac{\Omega^2 
    W}{L^2}\frac{B_{0}^2}{\Sigma} \left| 
    \Phi_{M}\right|^2}&\displaystyle{  \dds\ln\left(\frac{W\Sigma}{B_{0}^2}\right)\
    \biggr]_{corot}}
\end{eqnarray*}
or:
\begin{eqnarray}
    \displaystyle{\delta\omega}&\displaystyle{ \int_{s_{min}}^{s_{max}}
    ds\,\,\,
    2\frac{B_{0}^2}{\Sigma} \left|\dd {s} \Phi_{M}\right|^2
    \left(\frac{-2\tom}{L^2}\right) }\!\!\!\!\!\!&\nonumber\\
    &&\nonumber\\
    &\displaystyle{=-8i\pi\biggl[\frac{\Omega^2 
    W}{\Omega'L^2}\frac{B_{0}^2}{\Sigma} \left| 
    \Phi_{M}\right|^2 \dds\ln}&\displaystyle{\left(\frac{W\Sigma}{B_{0}^2}\right)
    \biggr]_{corot} }
\label{eqn:growthrate}
\end{eqnarray}
where the square bracket in the \rhs is evaluated at the 
pole.  The main contribution to the integral on the \lhs comes 
form the inner cavity, at $\tom<0$.  This integral is thus real and 
positive.  On the \rhs, $\Omega'$ is usually negative, so 
that we get the sign of the growth rate $\gamma={\rm Im}(\omega)$ 
resulting from the resonance:
\begin{equation}
    {\rm sign}(\gamma)={\rm sign}\left[\dds\ln\left(\frac{W\Sigma} 
    {B_{0}^2}\right)\right]_{corot}
    \label{eqn:sgngamma}
\end{equation}
Thus outward gradients of $W$ and $\Sigma$ are stabilizing, while an 
outward gradient of $B_{0}$ is destabilizing.  

The full variational principle given in Appendix A contains a second 
resonant term, associated with pressure stresses.  This term is 
proportional to the derivative of the specific vorticity (sometimes 
called {\it vortensity} in this context):
\begin{equation}
    {\rm sign}(\gamma)\biggr|_{\rm Pressure}= {\rm 
    sign}\left[\dds\ln\left(\frac{W}{\Sigma}\right)\right]
    \label{eqn:gammapress}
\end{equation}
There is also a cross term, proportional to $hB$. Thus when magnetic 
stresses are absent we recover the usual result (Papaloizou and 
Pringle, 1985; Narayan \etal, 1987) on the effect 
of the corotation resonance.  Note that in this case an outward gradient 
of $\Sigma$ is destabilizing.  The result in the general case is a 
combination of these two contributions.

However another result of the general variational form is that the 
pressure contribution is proportional to $\left|\sigma\right|^2$; on the 
other hand, and in spite of the parallel mentioned in section 
(\ref{subsec:pert}) between $\sigma$ and $B_{z}^D$, the magnetic 
contribution is rather proportional to $\left|\Phi_{M}\right|^2$, where 
$\Phi_{M}$ is the integral of 
$B_{z}^D$.  As mentioned in section (\ref{subsec:WKB}), $\sigma$ is 
evanescent in the forbidden band at corotation, making the effect of the 
resonance exponentially small in the pressure-driven case.  But although 
$B_{z}^D$ is also evanescent, $\Phi_{M}$ is not!

In order to understand this, let us imagine that the perturbed field 
$B_{z}^D$ is strictly zero in the forbidden band, so that $\Phi_{M}$ has 
its source only in the inner cavity where waves can propagate.  Then in 
the corotation region $\Phi_{M}$ will have the classical behavior of an 
$m$-polar potential in vacuum, varying as $r^{-m}$. This allows it to 
retain a sizable effect at the corotation radius, and thus to 
efficiently exchange energy with the Rossby vortex.

It would be of course difficult to find a simple explanation to the 
stabilizing or destabilizing effect of the corotation resonance, in this 
case as in the Papaloizou-Pringle instability.  The starting point would 
certainly be that, when the equilibrium field varies across the 
corotation radius, it results in a variation of the tangential force 
exerted on the disk by the magnetic potential $\Phi_{M}$, produced by 
fluctuations in the inner disk region.  This variation (or, in the 
opposite direction, that due to a gradient in the disk inertia) in turn 
results in differential azimuthal motions, at the source of the Rossby 
vortex.

We will not try to get, from the variational principle, an estimate of 
the resulting growth rate.  Rough estimates show that it should be of 
the order of $v_{A}/r\sim\beta^{-1/2}\Omega h/r$.  We will rather turn 
now to numerical solution of the equations to confirm our analysis and 
exhibit explicit solutions.

\section{Numerical study}
\label{sec:Num}
\subsection{The eigenvalue problem}
\label{subsec:modnum}
We solve numerically the set of Eqs.~(\ref{eqn:Eulu} - \ref{eqn:contB}), 
for the four unknowns $U,\, V,\, h$ and $B_{z}^D$, by turning it into an 
eigenvalue problem.  To do this we write the equations on a staggered 
grid, with velocities measured halfway between density grid points.

The magnetic potential appearing in the \rhs of the equations is 
expressed through Eq.~(\ref{eqn:PoissonB}).  Special care is needed for 
the Poisson Kernel, which is divergent at $x=1$.  We use a smoothing, 
replacing Eq.~(\ref{eqn:kernel}) by:\\
\begin{eqnarray}
\label{eqn:Kernsmooth}
K_{m}(x)=\frac{x}{\pi}\int_{0}^{\pi}\frac{d\vt\cos 
m\vt}{(x^2-2x\cos\vt+1+\epsilon^2)^{1/2}}
\end{eqnarray}
where $\epsilon$ is of the order of $c_{S}/r\Omega$, representing the 
effect of finite disk thickness; we have checked that varying $\epsilon$ 
does not affect the results presented here. Then, for a grid of $n_{G}$ 
points separated by $ds$ we define a vector $\xi$ of length $4n_{G}$, 
such that
\begin{eqnarray*}
   &\xi(4i)&=U\left(s=(i+\frac1 2)ds\right)\\
   &\xi(4i+1)&=V\left(s=(i+\frac1 2)ds\right)\\
   &\xi(4i+2)&=h\left(s=i\,ds\right)\\
   &\xi(4i+3)&=B_{z}^D\left(s=i\,ds\right)
\end{eqnarray*}
with $i=1$\ldots$\,n_{G}$.  In this manner our equations can be set as an 
eigenvalue problem:

\begin{equation}
    -i\omega\xi={\cal M}\,\xi
    \label{eqn:mat}
\end{equation}

where ${\cal M}$ is a ($4n_{G}$x$4n_{G}$) matrix.  It would be band 
diagonal if we were dealing with a differential problem, \ie local 
forces.  But the magnetic potential (\ie long range, non-local forces), 
which turns it into an integro-differential problem, makes ${\cal M}$ a 
full complex matrix, for which we seek eigenvalues and eigenvectors. ${\cal 
M}$ would be hermitian in the absence of the radiating boundary condition 
at large $s$, and of the corotation resonance.

One usually considers that 1000x1000 is the maximal size of a matrix 
that can be inverted numerically without excessive accumulation of 
roundoff errors.  This leaves us with a maximum $n_{G}=250$: in 
practice, using the complex integration path described below, we have 
found this limit largely sufficient to get very good convergence of 
eigenvalues and eigenvectors, and we have not tried to challenge it.  
The determination of eigenvalues and eigenvectors is thus done with 
standard library packages and takes only a few minutes on a modern 
workstation.

\subsection{The complex contour in $s$}
\label{subsec:complaxis}

The main limit to the precision comes from the vicinity of the 
corotation radius, where the solution varies rapidly--the more as we 
consider very weakly amplified solutions, so that the pole lies very 
close to the real-$s$ axis.  We solve this difficulty, and at the same 
time the outer boundary condition, by solving on an oblique axis in the 
complex-$s$ plane.

This means that, taking $s_{min}=0$ without loss of generality (since  
$s$ is logarithmic), we choose $s_{max}$ complex with a positive 
imaginary part.  This serves at the same time three purposes:
\begin{enumerate}
    \item It allows us to move the integration axis far from the pole, 
    in order to get a smoother behavior.  Given the limited number of 
    grid points we can use, this strongly improves the accuracy of the 
    eigenvalues.  On the other hand it allows us to exhibit the 
    eigenvector only on this complex axis, so that its components (the 
    velocities etc.)  are not the physical solutions one would obtain by 
    solving on the real axis.  However we maintain the obliquity of the 
    integration axis weak enough to limit this problem.  In selected 
    cases, where the growth rate is strong enough, we solve on the 
    real-$s$ axis and find, as could be expected, that the solution is 
    significantly modified only at large $s$ (note that our solution on 
    the complex axis {\it is} correct, but is not the one physically 
    observable in real space).

    \item Given the constraints on the inverse Laplace transform, 
    discussed in section \ref{subsec:variat}, solution along this 
    oblique axis allows us to correctly describe damped solutions by 
    analytic continuation, as long as the imaginary part of $s_{max}$ is 
    sufficient that the integration contour lies on the correct side.  
    We will see below that this appears in a very obvious manner in the 
    eigenvalue spectra.
    
    \item The boundary condition we apply at $s_{max}$ is the {\it 
    radiation condition}, also called the {\it outgoing wave condition}.  
    This condition states simply that we do not want the system to 
    receive information from outside, so that the only perturbation 
    allowed at $s_{max}$ is the wave traveling outward.  In our case, 
    far enough beyond corotation that the WKB condition is valid, this 
    translates into accepting only a trailing wave, with a local radial 
    wavenumber $k>0$.  We show in Appendix \ref{apdx.outBC} that this 
    same obliquity of the integration axis (with Im($s>0$)) provides a 
    simple and efficient way of implementing this boundary condition.  
\end{enumerate}
\begin{figure}
\psfig{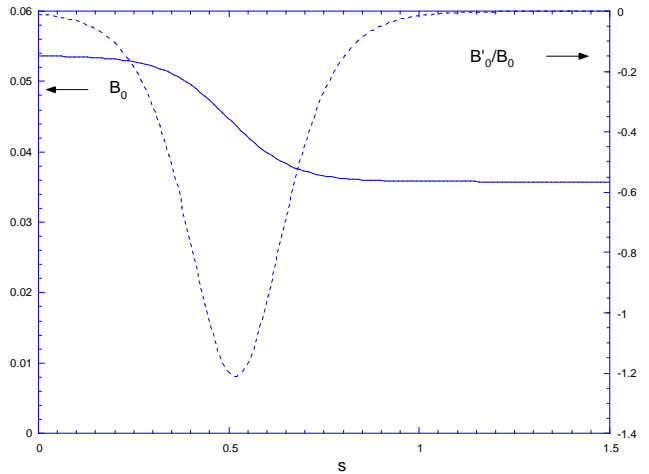}
\caption{\label {fig:profs} \scriptsize
A typical profile of the magnetic field (solid) and its logarithmic 
derivative (dots). We have a strong but localized gradient, resulting 
in a limited total change of $B_{0}$. A similar profile is used for the 
surface density of the disk}\end{figure}
\subsection{The profiles}
\label{subsec:profs}
We wish to describe an equilibrium with gradients of density and 
vertical magnetic fields.  On the other hand we do not wish too strong 
changes of these quantities, which would cause large differences in the 
radial wavelengths and thus create numerical problems.  Thus we use the 
analytical result, showing that amplification depends only on the 
local gradients at corotation.  This allows us to take profiles with 
strong but localized gradients, giving limited total variations across the 
integration domain:
\begin{eqnarray}    
    \Sigma(s)&=&\Sigma_{0}\left[1.-d_{\Sigma}\tanh\left(p_{\Sigma}(s-s_{0})\right)
    \right]\nonumber\\
   B_{0}(s)&=&B_{0int}\left[1.-d_{B}\tanh\left(p_{B}(s-s_{0})\right)
    \right]
    \label{eqn:profs}
\end{eqnarray}
Using $d_{\Sigma}$ or $d_{B}\sim .2$ gives reasonable profiles for our 
purpose.  This does not claim to represent realistic profiles, but 
allows us an optimal study of the physics involved.  
Fig.~(\ref{fig:profs}) shows a typical profile of the magnetic field and 
its logarithmic derivative.  When needed we use trial and errors, on 
low-precision runs with a limited number of points, to change $s_{0}$ so 
that the corotation radius lies close to the maximal gradient.

\subsection{Numerical results}
\label{subsec:resnum}
We measure the strength of the magnetic field by the usual parameter 
giving the ratio of thermal to magnetic pressure: 
\[\beta=\frac{8\pi 
p}{B_{0}^2}\]
taken at the inner disk radius.  At low field 
($\beta\gg 1$) we expect the magnetic component of the perturbation to 
play only a weak role.  Furthermore, this is the regime where the 
magneto-rotational instability is unstable and should dominate anomalous 
viscosity.  Nevertheless we will present numerical results at 
$\beta=50$, in order to check the general physics presented here.  In 
contrast we will show a second set of results for $\beta=1$, where our 
instability is obtained while the magneto-rotational instability (which 
does not appear in our analysis since we consider infinitely thin disks) 
should be stable.

\begin{figure}
\psfig{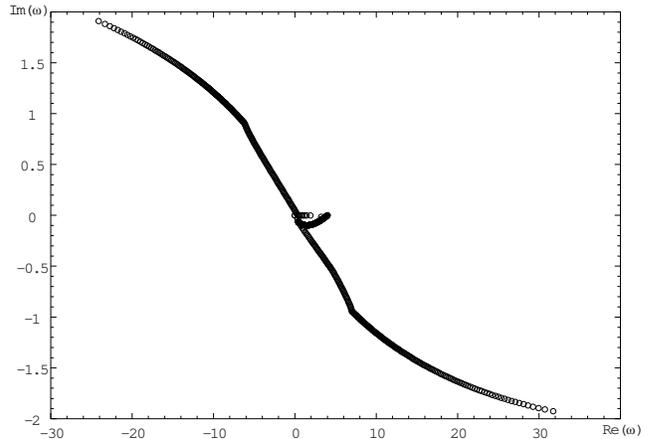}
\caption{\label {fig:spectrumb50} \scriptsize
The full spectrum of eigenvalues in a typical low-field case, with 
$\beta=50.$ and $m=4$. Integration is performed on an oblique complex 
axis $s=(0.,\ 0.)$ to $(1.5,\ .1)$. Most of the eigenvalues are 
associated to the discretization of the system.}\end{figure}

\begin{figure}
\psfig{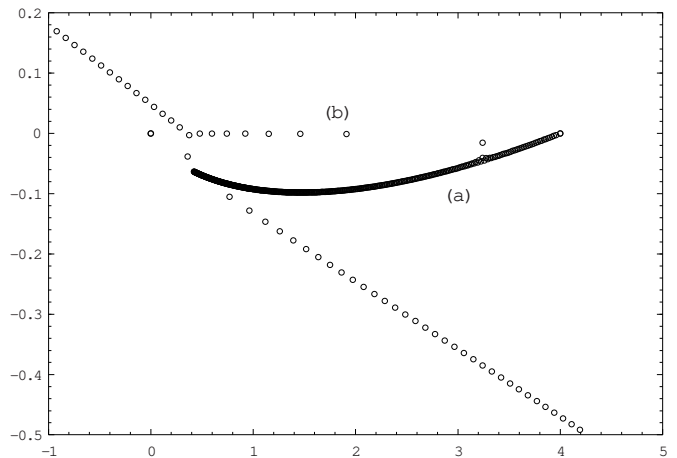} \caption{\label 
{fig:spectrumb50zoom} \scriptsize A zoom on the results of 
Fig.~\ref{fig:spectrumb50}.  Most noticeable are (a) a dense component 
corresponding to a continuous spectrum with corotation on the 
integration axis (b) another component corresponding to the physical 
discrete spectrum, close to Im($\omega$)=0.  }\end{figure}

We use the result of Eq.~(\ref{eqn:growthrate}) to limit the 
parameters we vary: we keep $\Sigma$ constant and introduce only a 
radial variation of $B_{0}$.  We use trial and error to have a maximum 
gradient very near the corotation radius, with \[\dd {s} \ln B_{0} 
\approx 1\] so that, from Eq.~(\ref{eqn:growthrate}), the modes 
should be weakly unstable if the magnetic contribution dominates, and 
(from the additional terms included in Eq.~(\ref{eqn:varprinct}) )
damped if pressure dominates.  We choose a sound speed $c_{S}=.1 
r_{0}\Omega_{0}$, where the subscript 0 notes values at the inner radius 
($s=0$).  Figs.~(\ref{fig:spectrumb50}--\ref{fig:spectrumb50zoom}) 
show the spectrum of eigenvalues we obtain for $\beta=50$ and $m=4$.  
Frequencies are normalized with $\Omega_{0}=1.$, so that eigenvalues 
whose real part lie between 0 and $m$ have a corotation at $s>0$.
\begin{figure}
\psfig{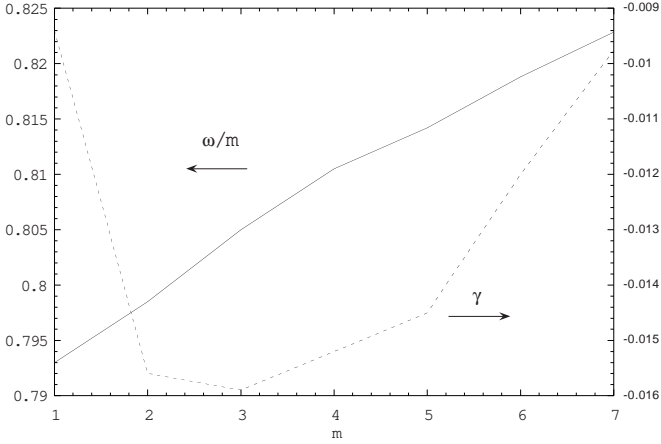} \caption{\label 
{fig:wb50} \scriptsize Real (full line) and imaginary (dashed) parts of 
the $n=0$ eigenvalue for $\beta=50$. All the modes are damped by the 
corotation resonance.
}\end{figure}
\begin{figure}
\psfig{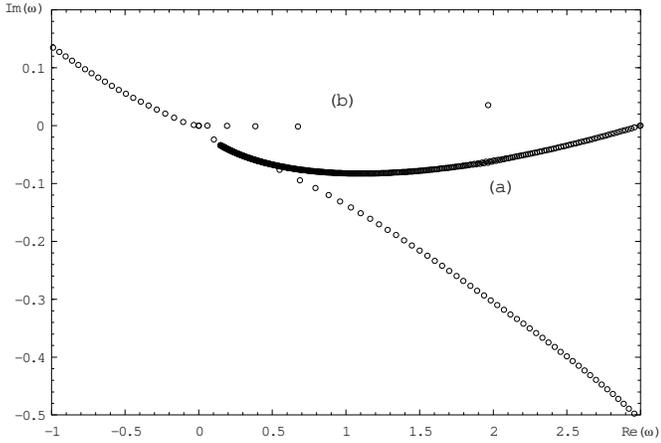} \caption{\label 
{fig:spectrumb1zoom} \scriptsize The spectrum for $\beta=1$, $m=3$. The 
fundamental mode has a significant positive real part, \ie is amplified 
by the corotation resonance.
}\end{figure}
\begin{figure}
\psfig{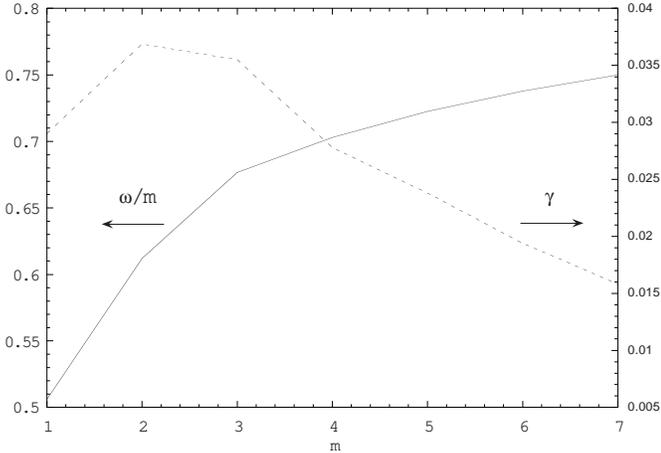} \caption{\label 
{fig:wb1} \scriptsize Real (full line) and imaginary (dashed) parts of 
the $n=0$ eigenvalue for $\beta=1$. The modes have become unstable by the 
corotation resonance.
}\end{figure}
\begin{figure}
\psfig{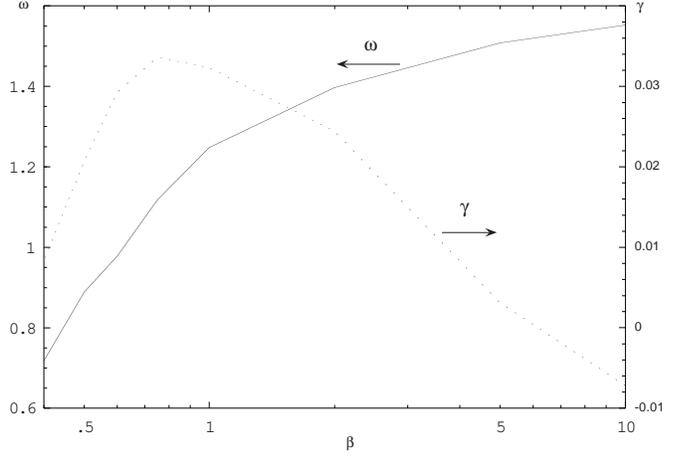} \caption{\label 
{fig:wm2} 
\scriptsize The frequency and growth rate as a function of $\beta$, for 
the $m=2$ mode.
}\end{figure}
\begin{figure}
\psfig{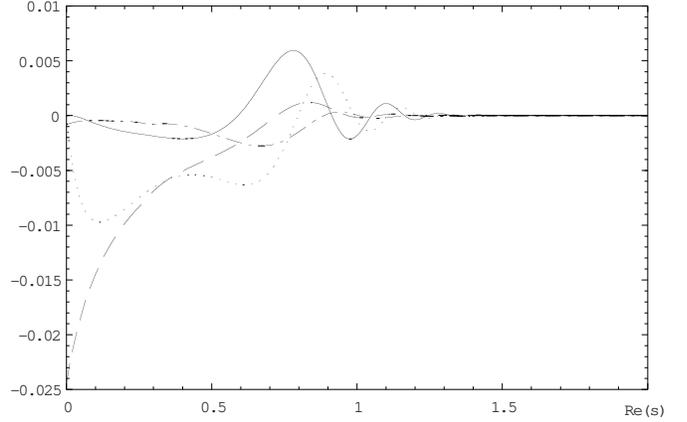} \caption{\label {fig:vpb1m3} 
\scriptsize The eigenvector: Real (solid) and imaginary (dots) parts of 
U, Real (dashes) and imaginary (dash-dots) parts of V for $\beta=1$, 
$m=3$. Corotation is at $s=.28$.
}\end{figure}
We distinguish three components in the spectrum:
\begin{enumerate}
    \item a set of values aligned diagonally, as seen in Fig.~(\ref{fig:spectrumb50}).  
    From their behavior as the number $n_{s}$ 
    of grid points is varied, we find these values to form a continuous 
    spectrum as $n_{s}\rightarrow\infty$, associated with the 
    discretization of the problem, and thus not physical. 
    
    \item A second component, labeled (a) in 
    Fig.~(\ref{fig:spectrumb50zoom}).  These also form a continuous 
    spectrum, and correspond to unphysical solutions with corotation 
    very near grid points: thus when we solve along a real-$s$ axis, 
    this component lies on the real-$\omega$ axis.  This permits us to 
    directly check on the plots that all roots above this set, even when 
    (as here) they are weakly damped, lie on the proper side of the 
    integration axis, as discussed in section \ref{subsec:complaxis}.

    \item A third component, labeled (b) in 
    Fig.~(\ref{fig:spectrumb50zoom}), corresponds to a discrete 
    spectrum.  In order of decreasing $\omega$, these eigenvalues 
    correspond to the classical set of modes with $n=0,\ 1,\ 2,\ldots$ 
    nodes in their radial structure, between the inner radius and 
    corotation.  Due to the limited numerical precision, the frequency 
    of only the fundamental solution, $n=0$, is reliable, although 
    higher-order ones do show the increasing number of nodes, verifying 
    our identification of this discrete spectrum.  The $n=0$ eigenvalue 
    converges to a very good accuracy when $n_{s}$ is varied above 
    typically 200.    
\end{enumerate}
\begin{figure}
\psfig{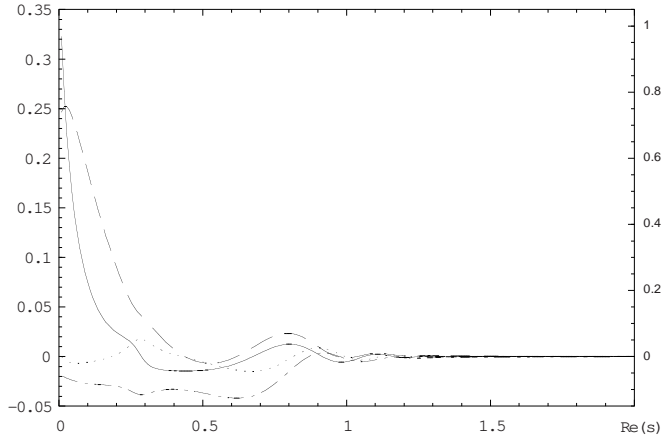} \caption{\label 
{fig:vpphib1m3} 
\scriptsize The eigenvector: Real (solid) and imaginary (dots) parts of 
$\sigma$, Real (dashes) and imaginary (dash-dots) parts of $\Phi_{M}$ for $\beta=1$, 
$m=3$. Corotation is at $s=.28$.
}\end{figure}
Fig.~(\ref{fig:wb50}) shows the variations of the pattern frequency, 
$\omega/m$, and growth rate, $\gamma={\rm Im}(\omega)$, as a function of 
$m$.  At this high value of $\beta$, the modes are all weakly damped by 
the corotation resonance, according to Eq.~(\ref{eqn:gammapress}).

On the other hand, Fig.~(\ref{fig:spectrumb1zoom}) shows the spectrum 
for $\beta=1.$, $m=3$. The fundamental mode has acquired a significant 
positive imaginary part, \ie becomes unstable by the corotation 
resonance. Fig.~(\ref{fig:wb1}) shows that low-$m$ modes become  
unstable. The growth rate is not strong (\eg for $m=2$, $\gamma$ is .06 
times the pattern frequency), but one has to remember that we are 
dealing with normal modes, \ie long-lived, exponentially amplified 
structures, so that this may result in strong amplitudes after a finite 
time.

We show on Fig.~(\ref{fig:wm2}) the mode frequency 
and growth rate for $m=2$, as a function of $\beta$. The frequency 
decreases as $\beta$ decreases, while the growth rate has 
a maximum for $\beta$ close to 1. This can be understood from the 
dispersion relation, Eq.~(\ref{eqn:reldisp}): at low $\beta$ the 
wavelength becomes larger, widening the ``cavity'' and thus pushing to 
larger $r$ the corotation radius ; on the other hand the turning point 
(the radius where the radial wavenumber $k_{s}=0$) is in fact not exactly 
at the Lindblad resonance, but where
\begin{equation}
\tom^2-\kappa^2=\frac{2B_{0}^2}{\Sigma}\frac m r + \frac{m^2}{r^2}c_{S}^2
\end{equation}
so that the forbidden band around corotation widens as $\beta$ decreases. 
Thus at low $\beta$ the amplification mechanism becomes less efficient.

Finally we show on Figs.~(\ref{fig:vpb1m3} - \ref{fig:vpphib1m3}) the 
eigenvector for a typical mode, showing in particular the cavity inside 
corotation and outgoing wave beyond it, decreasing at large $s$ because 
of the obliquity of the integration axis.
\section{Discussion}
\label{sec:disk}
We have presented an instability of magnetized accretion disks, which 
has the important potential of relating accretion (by a spiral density 
wave in the disk) and ejection, through Alfven waves propagating to the 
corona of the disk the energy and angular momentum extracted by the 
spiral. The instability has the following characteristics:

\begin{itemize}
    \item  It forms {\it normal modes}, \ie standing, exponentially 
    growing, wave patterns.

    \item Their growth rate is rather small, typically of the order of 
    $h\Omega/r$ where $h$ is the disk thickness (we could have shown 
    stronger growth rates by increasing the gradients in our numerical 
    examples). But their exponential behavior still allows them to reach 
    high amplitudes in a finite time.

    \item The modes are localized in the innermost part of the disk: 
    since the sound and Alfven velocities are small compared to the 
    rotation velocity, the wavelength in the ``cavity'', between the 
    inner radius and the ILR, is of the order of a few times the disk 
    thickness (we will not discuss here the exact relationship, which is 
    not important).  Thus the frequency of the fundamental mode (with no 
    node in the cavity) is essentially fixed, at a given $m$, by the 
    existence of an ILR close to the inner disk radius: 
    $\omega-m\Omega=-\kappa$ at $r=r_{i}$ close to the inner radius 
    $r_{0}$, with $\Omega=\kappa$ in a keplerian disk, gives 
    $\omega=m\Omega(r_{c})=(m-1)\Omega(r_{i})$, where $r_{c}$ is the 
    corotation radius, and thus 
    \[r_{c}=\left(\frac{m}{m-1}\right)^{2/3}r_{i}\]
    
    Thus these modes cannot be considered as candidates to explain 
    accretion in the bulk of the disk, but only to energize jets close 
    to the inner radius, where, in usual viscous models, half of the 
    energy and momentum of accretion are deposited in a boundary layer. 
    Actually the size of the ``cavity'' is such that it might be 
    considered as a form of boundary layer.
    
    \item  The instability occurs in strongly magnetized disks, with 
    $\beta\simlt 1$ (where the magneto-rotational instability is stable)

    \item It appears when the quantity ${\Omega \Sigma}/{B_{0}^2}$ 
    increases radially.  This requires either an inverted density 
    gradient, or a strong magnetic field gradient.  This means that disk 
    models must consider the radial transport of magnetic flux, as well 
    as that of the accreted matter: the horizontal part of the magnetic 
    flux can be expelled from the disk vertically by buoyancy (\eg the 
    Parker instability, Foglizzo and Tagger, 1994, 1995), but the 
    vertical part can only accumulate in the central region of the disk, 
    if the accretion turbulence transports it together with the matter.  
    One may note as an example that at the center of our galaxy, on a 
    scale of $\sim 100$ pc, the magnetic field seems to be vertical and 
    of the order of a few milligauss, as compared with a horizontal 
    field of a few $\mu$gauss elsewhere in the galactic disk.

\end{itemize}

This threshold in radial gradients also implies that this instability is 
very likely to result in a bursty behavior of the accretion/ejection 
process, associated with the processing of both gas {\it and magnetic flux} 
in the disk.\\

We also wish to mention the possible connection between this 
instability and the QPOs observed in X-ray binaries: since the mechanism 
favors low $m$ values, \ie spirals with a small number of arms, it is 
very likely to appear as very coherent, monochromatic structures, in the 
manner of galactic spirals.  Their corotation radius would be of the 
order of a few times the inner disk radius. These are precisely 
characteristics of the low frequency QPOs, with a frequency in the 
range of .1 (in black hole binaries) to a few tens (in neutron star
binaries) of Hertz (see \eg Wijnands and Van der Klis, 1999), \ie a 
frequency typical of keplerian rotation not far from the inner disk 
boundary (Markwardt \etal, 1999 ; Psaltis \etal, 1999). 

\acknowledgements The authors gratefully acknowledge many helpful 
discussions with T.~Foglizzo, J.~Goodman, R.N.~Henriksen and J.~Papaloizou.
Part of this work was completed at the Isaac Newton Institute for 
Mathematical Sciences (Cambridge) while one of us (MT) attended the 
program on Astrophysical Disks.

\appendix
\section{The full variational principle}
\label{apdx.appvar}
In this appendix we perform essentially the same computations as in 
section (\ref{subsec:variat}), but retaining pressure stresses to give 
a complete variational principle. 

Thus Eq.~(\ref{eqn:Bzd}) becomes:
\begin{eqnarray}
& r^2B_{z}^D=-\displaystyle{\dd {s} \,\left( \frac{2}{L}\frac{B_{0}^2} 
{\Sigma}\, \dd {s} \Phi_{M}\right)}-\displaystyle{\dd {s} \,\left( 
\frac{B_{0}} {L}\, \dd {s} \cs2 h\right)}\nonumber 
\\
 &\nonumber\\
    &\displaystyle{+\frac{2B_{0}^2}{\Sigma L^2}\Phi_{M} 
    \biggl\{m^2\left(L+4\Omega\Omega'\right)}\ \ \ \ \ \ \ \ \ \ \ \ \ 
    \ \ \ \ \ \ \ \ \ \ \ \ \ \ \ \ \ \ \ \ \ \ \ \ \ \ \nonumber \\
 &\nonumber\\
   & \!\!\!\!\!\!\!\!\!\!\!\!\!\!\!\!\displaystyle{+2m\Omega\biggl[ 
    \displaystyle{\tom\dd {s} \ln\left(\frac{\Omega 
    B_{0}^2}{\Sigma}\right)}+2\frac{\Omega 
    W}{\tom}\dd {s} \ln
    \left(\frac{W\Sigma}{B_{0}^2}\right)\ \biggr]\ \biggr\}}\nonumber \\
 &\nonumber\\
    &\displaystyle{+\frac{B_{0}}{L^2}\cs2 h 
    \biggl\{m^2\left(L+4\Omega\Omega'\right)} \ \ \ \ \ \ \ \ \ \ \ \ \ 
    \ \ \ \ \ \ \ \ \ \ \ \ \ \ \ \ \ \ \ \ \ \ \ \ \ \ \nonumber \\
 &\nonumber\\
    &\!\!\!\!\!\!\!\!\!\!\!\!\!\!\!\!\displaystyle{+2m\Omega\biggl[ 
    \displaystyle{\tom\dds \ln\left(\Omega 
    B_{0}\right)}+2\frac{\Omega 
    W}{\tom}\dd {s} \ln
    \left(\frac{W}{B_{0}}\right)\ \biggr]\ \biggr\}}
\label{eqn:Bzdt}
\end{eqnarray}

In the same manner, starting from Eq.~(\ref{eqn:contin}) we obtain 
a similar equation for $\sigma$:
\begin{eqnarray}
&r^2\sigma=-\displaystyle{\dd {s} \,\left( \frac{2B_{0}}{L}\, \dd {s} 
\Phi_{M}\right)}-\displaystyle{\dd {s} \,\left( \frac{\Sigma} {L}\, \dd 
{s} \cs2 h\right)}\ \ \ \ \ \ \ \ \ \ \ \ \ \nonumber \\
 &\nonumber\\
    &\displaystyle{+\frac{2B_{0}}{L^2}\Phi_{M} 
    \biggl\{m^2\left(L+4\Omega\Omega'\right)}\ \ \ \ \ \ \ \ \ \ \ \ \ 
    \ \ \ \ \ \ \ \ \ \ \ \ \ \ \ \ \ \ \ \ \ \ \ \ \ \ \nonumber \\
 &\nonumber\\
   & \!\!\!\!\!\!\!\!\!\!\!\!\!\!\!\!\displaystyle{+2m\Omega\biggl[ 
    \displaystyle{\tom\dd {s} \ln\left(\Omega 
    B_{0}\right)}+2\frac{\Omega 
    W}{\tom}\dd {s} \ln
    \left(\frac{W}{B_{0}}\right)\ \biggr]\ \biggr\}}\nonumber \\
 &\nonumber\\
    &\displaystyle{+\frac{\Sigma}{L^2}\cs2 h 
    \biggl\{m^2\left(L+4\Omega\Omega'\right)} \ \ \ \ \ \ \ \ \ \ \ \ \ 
    \ \ \ \ \ \ \ \ \ \ \ \ \ \ \ \ \ \ \ \ \ \ \ \ \ \ \nonumber \\
 &\nonumber\\
    &\!\!\!\!\!\!\!\!\!\!\!\!\!\!\!\!\displaystyle{+2m\Omega\biggl[ 
    \displaystyle{\tom\dds \ln\left(\Omega 
    \Sigma\right)}+2\frac{\Omega 
    W}{\tom}\dd {s} \ln
    \left(\frac{W}{\Sigma}\right)\ \biggr]\ \biggr\}}
\label{eqn:ht}
\end{eqnarray}

We multiply Eq.~(\ref{eqn:ht}) by $\cs2 h^*$, and Eq.~(\ref{eqn:Bzdt}) 
by $2\Phi_{M}^*$, sum the results and integrate over $s$: this gives, 
after integrations by parts, the full variational principle similar to 
Eq.~(\ref{eqn:varprinc}):\\
\begin{eqnarray*}
&\displaystyle{\int_{s_{min}}^{s_{max}}ds\, \biggl\{r^2(s)}&\displaystyle{ 
\!\!\!\!\!\!\left[\cs2 h^*\sigma + 2\Phi_{M}^*(s)B_{z}^{D}(s)\right]} \\
& &\\
&-\displaystyle{\frac{2B_{0}\cs2}{L}\biggl[\left(\dds h^*\right)}
&\displaystyle{\left(\dds\Phi_{M}\right)+\left(\dds\Phi_{M}^*\right)\left(\dds 
h\right) \biggr]}\\
& &\\
&-\displaystyle{\frac{1}{L}\biggl[\Sigma c_{S}^4\left|\dds h\right|^2+}
&\displaystyle{ 
4\frac{B_{0}^2}{\Sigma}\left|\dds \Phi_{M}\right|^2\biggr]}\\
& &\\
&-\displaystyle{\frac{2B_{0}}{L^2}\cs2
\biggl[h^*\Phi_{M}+}
&\displaystyle{h\Phi_{M}^*\biggr]
\biggl\{m^2\left(L+4\Omega\Omega'\right)} \\
& &\\
& \displaystyle{\,\,\,\,\,\,\,\,\,\,\,\,\,\,\,\,\,\,+2m\Omega 
\biggl[\tom} &\displaystyle{ \dds\ln\left(\Omega 
B_{0}\right)+2\frac{\Omega W}{\tom}\dds \ln 
\left(\frac{W}{B_{0}}\right)\ \biggr]\ \biggr\} }\\
& &\\
&+\displaystyle{\frac{\Sigma}{L^2}c_{S}^4\left| h\right|^2\biggl\{m^2}
&\displaystyle{
 \left(L+4\Omega\Omega'\right)}\\
& &\\
& \displaystyle{\,\,\,\,\,\,\,\,\,\,\,\,\,\,\,\,\,\,+2m\Omega 
\biggl[\tom} &\displaystyle{ \dds\ln\left(\Omega 
\Sigma\right)+2\frac{\Omega W}{\tom}\dds \ln 
\left(\frac{W}{\Sigma}\right)\ \biggr]\ \biggr\} }\\
& &\\
&+\displaystyle{\frac{4B_{0}^2}{\Sigma L^2}\left| \Phi_{M}\right|^2
\biggl\{ }
&\displaystyle{\!\!\!\!m^2\left(L+4\Omega\Omega'\right)}\\
& &\\
& \displaystyle{+2m\Omega\biggl[ \tom\dds 
\ln} 
&\displaystyle{\!\!\!\!\!\!\!\!\left(\frac{\Omega B_{0}^2}{\Sigma}\right)+2\frac{\Omega W}{\tom}\dds 
\ln \left(\frac{W\Sigma}{B_{0}^2}\right)\ \biggr]\ \biggr\}\ 
\biggr\}}\\
\end{eqnarray*}
\begin{eqnarray}
    &=\displaystyle{-\biggl[\frac{2B_{0}}{L}\cs2\left(h^*\dds\Phi_{M}+\Phi_{M}^*\dds 
    h\right)} \nonumber\\
    &\,\,\,\,\,\,\,\,\,\,\,\,\,\,\,\,\,\,\,\,\,\,\,\,\,\,\,\,\,\,\, 
    \displaystyle{+\ \frac{\Sigma}{L}c_{S}^4 h^*\dds 
    h\ +\ \frac{4B_{0}^2}{\Sigma 
    L}\Phi_{M}^*\dds\Phi_{M}\biggr]_{s_{min}}^{s_{max}}}
\label{eqn:varprinct}
\end{eqnarray}

As in Eq.~(\ref{eqn:varprinc}) the \lhs is hermitian, so 
that this is again a variational principle.

\section{Boundary conditions}
\label{apdx.BC}

\subsection{Outer boundary: outgoing wave}
\label{apdx.outBC}
As explained in section \ref{subsec:complaxis}, we require that at the outer 
boundary the solution reduces to an outgoing wave.  The complex 
integration axis allows us an easy implementation of this condition.  We 
start by assuming that $s_{max}$ is large enough that the WKB 
approximation can be used. 

Then let us consider the two WKB solutions: \[\xi_{\pm}\sim 
\exp\left(\pm i\int^s ds'k(s')\right)\] where the plus and minus sign 
correspond respectively to the trailing (outgoing) and leading 
(incoming) waves.  If we integrate on an oblique axis in the complex-$s$ 
plane, with Im($s$) positive, one sees that the trailing solution 
decreases exponentially, while the leading one grows 
exponentially, as $s$ increases.
    
Let us now assume that we have at $s_{max}$ an incorrect boundary 
condition.  In general, this means that we have leading and trailing 
waves of comparable amplitude.  But as one 
goes inward from $s_{max}$, the trailing solution grows exponentially as 
$s$ decreases while the leading one decreases.  Thus at some distance 
inward from $s_{max}$ the correct (trailing) solution will dominate.  
This method of dealing with the boundary condition has been introduced 
in disk physics by Drury (1980).  It allows us here to take any simple 
(but a priori false) boundary condition at $s_{max}$, and only check a 
posteriori that the correct behavior results; if it is not the case we 
just need to increase Im($s_{max}$).  In practice we use a 
reflecting condition at $s_{max}$ and have little difficulty with the 
outer boundary condition .

In fact we have another effect helping us with the outer boundary 
condition: as seen from the WKB dispersion relation, 
Eq.~(\ref{eqn:reldisp}), the wavelength becomes small at large $s$.  On 
the other hand our numerical scheme for the discretization of the 
equations is such that, when the wavelength becomes comparable with the 
grid spacing, waves suffer a heavy numerical damping.  Then if we have a 
solution resulting in an outgoing wave, this wave is damped at large $s$ 
and very little of its energy can be reflected.  Thus taking 
Re($s_{max}$) large enough is also a good way of implementing the outer 
boundary condition.

Let us now turn to the contribution of the outer boundary to the 
variational principle, Eq.~(\ref{eqn:varprinc}). It is easily 
obtained in the WKB approximation, 
\[\dds \Phi_{M}=ik\Phi_{M}\]
with $k$ given by the dispersion relation, Eq.~(\ref{eqn:reldisp}).  
Let us treat this boundary term by perturbations, as we do for the 
resonance term: we see that, since it is imaginary, its contribution 
will be to the growth rate of the mode.  In fact one easily checks that 
with $k>0$ (for an outgoing wave) it is always destabilizing, as 
mentioned in section (\ref{subsec:WKB}).  This is the amplification 
mechanism discussed, in the shearing sheet approximation, in paper~I. 
The resulting growth rate is always quite small.
%

\subsection{Inner boundary: reflection}
\label{apdx.inBC}
Any condition at the inner edge of the disk, which does not change the 
energy of the wave, will reflect it with an efficiency of 1.  This is 
the classical picture leading to normal modes.  Various conditions 
satisfy this criterion, and we discuss them here with the help of the 
variational principle.  For simplicity we will discuss here only the 
pressureless variational principle of Eq.~(\ref{eqn:varprinc}), but 
the generalization to the complete problem is straightforward.

From Eq.~(\ref{eqn:varprinc}) we see that by making either 
\begin{equation}
    \Phi_{M}=0
    \label{eqn:bound1}
\end{equation}
or 
\begin{equation}
    \dds \Phi_{M}=0
    \label{eqn:bound2}
\end{equation}
the boundary term vanishes, so that the mode energy is not changed.  
Either of these conditions is thus an acceptable reflecting boundary 
condition.

On the other hand it is more usual to assume a rigid boundary and 
require that 
\begin{equation}
    U=0
    \label{eqn:bound3}
\end{equation}
at the inner edge.  In this case Eqs.~(\ref{eqn:Eulu}-\ref{eqn:Eulv}), 
forgetting the pressure terms, give:
\begin{equation}
    4im\frac{B_{0}^2}{\Sigma^2}\Phi_{m}^*\dds \Phi_{M}
    =2i\tom^*\Omega \left|V\right|^2
\end{equation}
making the boundary term real.  Thus if we start from a marginal mode 
($\omega$ real) with boundary conditions (\ref{eqn:bound1}) or 
(\ref{eqn:bound2}), and change it to $U=0$, the resulting change to 
$\omega$ is real: this means that condition (\ref{eqn:bound3}) also 
gives a perfect reflection of the wave, but with a different phase; the 
resulting change in the integral phase condition modifies the discrete 
set of mode frequencies, but does not introduce growth or damping.  We 
note that the condition $V=0$ used by NGG would give a similar result.  
We use condition (\ref{eqn:bound3}) in the numerical solutions.

\section{Emission of Alfven waves in the WKB limit}
\label{apdx.alfven}
In this appendix we describe in the simplest manner the emission of
Alfven waves, carrying some of the energy and angular momentum of the
density waves in the disk to its corona if the latter has a small but
non-vanishing density (thus relaxing the hypothesis, made throughout the
paper, that the disk lies in vacuum).  We do this in a simple WKB limit
in the region where density waves propagate radially, ie far from the
corotation region.  A full computation, without the WKB approximation,
would be necessary to describe the emission of energy and momentum from
the vortex at corotation.  We defer it to a future publication.

In order to describe the vertical propagation of the Alfven waves, we 
now consider the $z$ dendence of the variables. 
We start from the Euler equations and 
derive equations for two new quantities:
\begin{eqnarray*}
    D(s,z) & = & \frac{\partial}{\partial s}U+imV=r^2\vec\nabla_{\perp}\cdot\vec 
    V_{\perp}\\
    R(s,z) & = & \frac{\partial}{\partial
    s}V-imU=r^2\vec\nabla_{\perp}\times\vec V_{\perp}\\
\end{eqnarray*}
where the subscript $\perp$ refers to the components of a vector in the 
plane of the disk. We get:
\begin{eqnarray}
     -i\tomega D  -2\Omega R- 2\Omega'(v-imu)&= r^2\vec\nabla\cdot\left(
     {\vec F}/{\rho_{0}}\right)\hfill
    \label{eqn:eulD}  \\
     -i\tomega R +WD+W'u &= r^2\vec\nabla\times\left({\vec
     F}/{\rho_{0}}\right)\hfill
    \label{eqn:eulR} 
\end{eqnarray}
where $\rho_{0}(s,z)$ is the equilibrium density, and $\vec F$ is the
perturbed force acting on the fluid.  Here for simplicity we will again
neglect pressure forces, unimportant for Alfven waves. 

We now use the WKB approximation, by defining a local radial wavenumber 
\[\frac{\partial}{\partial s} \simeq i k\]
with $k\gg m$. The dispersion relation, eq. (\ref{eqn:reldisp}), tells 
us that this is verified far from corotation. Consistently neglecting 
all radial gradients of equilibrium quantities, and using the induction 
equation for the perturbed magnetic field, we get after some algebra:
\begin{eqnarray}
     -i\tomega D -2\Omega R&=&
     \frac{v_{A}^2}{i\tomega}(\frac{q^2}{r^2}-\frac{\partial^2}{\partial
     z^2})D\hfill \label{eqn:WKBD} \\
     -i\tomega R +WD&=&
     -\frac{v_{A}^2}{i\tomega}\frac{\partial^2}{\partial z^2}R\hfill
     \hfill
    \label{eqn:WKBR} 
\end{eqnarray}

We now use the same procedure as in Tagger \etal (1992): let us assume
that at large $z$ ($z\gg h$, where $h$ is the disk scale height) the
density goes to a small but non-vanishing density $\rho_{\infty}$ 
mimicking a disk corona. 
There $v_{A}$ becomes very large, so that equation 
(\ref{eqn:WKBD}) gives to lowest order:
\[ D(z)\sim e^{-|kz|}\] where the minus sign comes from the condition
that $D$ decreases at infinity.  Thus we get the same behavior derived
for $\Phi_{M}$ in the case where the disk lies in vacuum.  On the other
hand equation(\ref{eqn:WKBR}) gives: 
\[ R(z)\sim e^{ik_{z}z}\]
where
\[k_{z}=\frac{\tomega}{v_{A}^\infty}, \ v_{A}^{\infty
2}=\frac{B_{0}^2}{4\pi\rho_{\infty}}\]
and causality requires that $k_{z}$ has the same sign as $\tomega$, so
that this describes an Alfven wave emitted from the disk.

We now multiply equations (\ref{eqn:WKBD}-\ref{eqn:WKBR}) by $\rho(z)$ 
and integrate over $z$. We get:
\begin{eqnarray}
     -i\tomega \bar D -2\Omega \bar R&=&
     \frac{\overline{v_{A}^2}}{i\tomega}\,\frac{q}{rh}\,\bar D\hfill
     \label{eqn:barD} \\
     -i\tomega \bar R +W\bar D&=&
     -\frac{\overline{v_{A}^2}}{i\tomega h}\, ik_{z}\,\bar R\hfill
     \hfill
    \label{eqn:barR} 
\end{eqnarray}
where the bar means an average over the disk height, and the \rhs of 
equation (\ref{eqn:barD}) is an estimate of the integral, which allows us 
to recover in the limit $\rho_{\infty}\rightarrow 0$ the dispersion 
relation, equation (\ref{eqn:reldisp}). From these equations we get the 
dispersion relation when the coronal density is small:
\begin{equation}
    \tomega^2=\kappa^2+\frac{|k|}{rh}\overline{v_{A}^2}-i\frac{k_{z}}{\tomega^2
    h}\kappa^2\overline{v_{A}^2}
\end{equation}
Using the value of $k_{z}$, and solving by perturbation, we find that
the emission of Alfven waves to the corona results in a damping of the
density wave, \ie $\omega$ gets a negative imaginary part:
\begin{equation}
    \gamma_{Alfven}\simeq -\frac{1}{2\tomega^2\beta^{1/2}}\kappa^2\Omega
\left(\frac{\rho_{\infty}}{\rho_{0}}\right)^{1/2}
\label{eqn:alfvendamping}
\end{equation}
where $\rho_{0}$ is the density at the midplane of the disk. From this 
result we can get the following conclusions:
\begin{itemize}
    \item  The damping is proportional to the square root of the density 
    at infinity, a result typical of magnetic breaking. 

    \item  It applies through $R$, the torsional component of the 
    perturbation: as expected, $R$ involves a torsional (in contrast 
    with $D$, giving a compressional) motion of the footpoints of the 
    field lines in the disk. This motion propagates to the corona as an 
    Alfven wave. It is noteworthy that $R$ is singular at corotation, 
    because of the vorticity gradient contribution ($W'u$) in 
    equation (\ref{eqn:eulR}), and indeed represents the Rossby vortex 
    there - although this is not described in the WKB 
    approximation used in this appendix. Thus we can expect a strong 
    Alfven wave emitted from thee Rossby vortex at corotation.

    \item  This may be even stronger, since equation 
    (\ref{eqn:alfvendamping}) shows a vanishing denominator at corotation. 
    This must of course be taken only as an indication, since the WKB 
    approximation vanishes in this region.
\end{itemize}
The full computation of this effect would need to relax the WKB 
approximation. Its consequences on the formation of a jet requires  
non-linear study of the deposition of Alfven wave energy in the corona. 
These extensions of our result will be the object of future work.


\end{document}